\documentclass[12pt]{article}
\usepackage{epsfig}
\usepackage{graphicx}
\usepackage{amssymb}
\usepackage{amsmath}
\usepackage{amsfonts}
\usepackage{mathrsfs}
\usepackage[dvips]{color}

\newcommand{\R}{\mathbb{R}}
\newcommand{\C}{\mathbb{C}}
\newcommand{\Z}{\mathbb{Z}}

\newcommand{\fa}{\mathfrak{a}}
\newcommand{\fb}{\mathfrak{b}}
\newcommand{\fc}{\mathfrak{c}}

\newcommand{\ff}{\mathfrak{f}}
\newcommand{\fg}{\mathfrak{g}}
\newcommand{\fh}{\mathfrak{h}}

\newcommand{\fu}{\mathfrak{u}}

\newcommand{\fn}{\mathfrak{n}}

\newcommand{\fD}{\mathfrak{D}}

\newcommand{\cH}{\mathcal{H}}

\newcommand{\be}{\begin{equation}}
\newcommand{\ee}{\end{equation}}
\newcommand{\bea}{\begin{eqnarray}}
\newcommand{\eea}{\end{eqnarray}}
\newcommand{\nn}{\nonumber}

\newcommand{\ed}{\end{document}}

\newcommand{\pbr}{\prec}
\newcommand{\pkt}{\succ}

\newcommand{\np}{\newpage}

\newcommand{\bi}{\begin{itemize}}
\newcommand{\ei}{\end{itemize}}

\newcommand{\bce}{\begin{center}}
\newcommand{\ece}{\end{center}}

\newcommand{\IM}{\,{\rm Im}}

\newcommand{\One}{{\stackrel{\leftrightarrow}{1}}}
\newcommand{\Zero}{{\stackrel{\leftrightarrow}{0}}}
\newcommand{\Ep}{{\stackrel{\leftrightarrow}{\mbox{\large{$\varepsilon$}}}}}
\newcommand{\ep}{{\stackrel{\;\leftrightarrow_\prime}{\mbox{\large{$\varepsilon$}}}}}
\newcommand{\MU}{{\stackrel{\leftrightarrow}{\mbox{\large{$\mu$}}}}}
\newcommand{\Mu}{{\stackrel{\;\leftrightarrow_\prime}{\mbox{\large{$\mu$}}}}}
\newcommand{\La}{{\stackrel{\;\leftrightarrow}{\mbox{\large{$\Lambda$}}}}}

\newcommand{\sF}{\mathscr{F}}
\newcommand{\sG}{\mathscr{G}}

\begin{document}

\title{Pseudo-Hermiticity and Electromagnetic Wave Propagation: The case of
anisotropic and lossy media}

\author{Ali~Mostafazadeh\thanks{Corresponding author, E-mail address:
amostafazadeh@ku.edu.tr, Phone: +90 212 338 1462, Fax: +90 212 338
1559} ~and~Giuseppe Scolarici\thanks{E-mail address:
giuseppe.scolarici@le.infn.it}
\\
Department of Mathematics, Ko\c{c} University, \\ 34450 Sar{\i}yer,
Istanbul, Turkey}

\date{ }
\maketitle

\begin{abstract}

Pseudo-Hermitian operators can be used in modeling electromagnetic
wave propagation in stationary lossless media. We extend this method
to a class of non-dispersive anisotropic media that may display loss
or gain. We explore three concrete models to demonstrate the utility
of our general results and reveal the physical meaning of
pseudo-Hermiticity and quasi-Hermiticity of the relevant wave
operator. In particular, we consider a uniaxial model where this
operator is not diagonalizable. This implies left-handedness of the
medium in the sense that only clockwise circularly polarized
plane-wave solutions are bounded functions of time.

\vspace{2mm}

\vspace{2mm}

\end{abstract}
\vspace{5mm}

\section{Introduction}

Consider a stationary (time-independent) dispersionless and
source-free medium with permittivity and permeability tensors
$\Ep=\Ep(\vec x)$ and $\MU=\MU(\vec x)$. Then the Maxwell's
equations are given by \cite{jackson}:
    \bea
    &&\vec\nabla\cdot(\Ep\vec E)=\vec\nabla\cdot\vec B=0,
    \label{max-1}\\
    &&\dot{\vec E}=\Ep^{-1}\fD \MU^{-1}\vec B,
    \label{max-E}\\
    &&\dot{\vec B}=-\fD\vec E,
    \label{max-B}
    \eea
where an overdot stands for a time-derivative, and $\fD$ is the curl
operator, e.g., $\fD \vec E :=\vec\nabla\times\vec E$. As shown in
Ref.~\cite{epl-2008}, Eqs.~(\ref{max-E}) and (\ref{max-B}) are
equivalent to
    \bea
    &&\vec B(\vec r,t)=\vec B_0(\vec r)-\int_0^t\fD\vec E(\vec
    r,\tau)\,d\tau,
    \label{B}\\
    &&\ddot{\vec E}+\Omega^2\vec E=0,
    \label{wave-eq}
    \eea
where $\vec B_0(\vec r):=\vec B(\vec r,0)$ and
    \be
    \Omega^2:=\Ep^{-1}\fD\,\MU^{-1}\fD.
    \label{Omega1}
    \ee
We can view $\fD$ as a linear operator acting in the Hilbert space
$\cH$ of square-integrable complex vector fields $\vec F:\R^3\to\C$.
This is defined by the $L^2$-inner product,
    \[\pbr\vec F|\vec G\pkt:=\int_{\R^3}\vec F(\vec x)^*\cdot\vec
    G(\vec x)~d\vec x^3,\]
where a dot placed between $\vec F^*$ and $\vec G$ stands for the
usual dot product.\footnote{For all $\vec F\in\cH$ , $\pbr\vec
F|\vec F\pkt<\infty$. This in particular implies that $|\vec F(\vec
x)|\to 0$ as $|\vec x|\to\infty$.}

It is easy to see that $\fD$ is a Hermitian operator acting in
$\cH$. Furthermore, for the cases that $\Ep$ and $\MU$ are given by
invertible Hermitian matrices, they also define invertible Hermitian
operators acting in $\cH$. This in turn implies that $\Omega^2$
satisfies the pseudo-Hermiticity relation \cite{p1}:
    \be
    \Omega^{2\dagger}=\Ep\,\Omega^2\Ep^{-1}.
    \label{ph}
    \ee
In particular, for a lossless (and gainless) medium $\Ep$ is a
positive-definite operator, and $\Omega^2$ is quasi-Hermitian
\cite{quasi}, i.e., it is related to a Hermitian operator via a
similarity transformation \cite{p2-p3}. This observation suggests an
interesting spectral method of solving the wave
equation~(\ref{wave-eq}). Ref.~\cite{epl-2008} gives a rather
explicit application of this method for an effectively
one-dimensional model involving localized inhomogeneity.
Ref.~\cite{p93} studies the possibility of extending this method to
a class of lossless dispersive media.

The purpose of the present article is to use the spectral properties
of the operator $\Omega^2$ in modeling electromagnetic wave
propagation in stationary non-dispersive anisotropic media that
display loss or gain. This requires considering non-Hermitian
permittivity and permeability tensors \cite{chew,someda}. In this
case, $\Omega^2$ is generally no longer quasi-Hermitian, and a
direct application of the spectral method outlined in
Ref.~\cite{epl-2008} is intractable. In the following we will
examine the structure of $\Omega^2$ for a lossy anisotropic medium.
Our investigation is motivated by the well-known fact that in many
physical applications one deals with such materials. A typical
example is a magnetic photonic crystals \cite{figotin}.\footnote{See
\cite{AG} for a comprehensive discussion of lossy crystals.}

\section{Solution of the Wave Equation}

The wave equation~(\ref{wave-eq}) admits the following formal
solution \cite{epl-2008}
    \be
    \vec E(\vec r,t)=\cos(\Omega t)\vec E_0(\vec r)+
    \Omega^{-1}\sin(\Omega t)\dot{\vec E}_0(\vec r),
    \label{E=1}
    \ee
where $\vec E_0(\vec r):=\vec E(\vec r,0)$, $\dot{\vec E}_0(\vec
x):=\dot{\vec E}(\vec r,0)$, and
    \be
    \cos(\Omega t):=\sum_{n=0}^\infty
    \frac{(-1)^n}{(2n)!}\;(t^2\Omega^2)^n,
    ~~~~~~\Omega^{-1}\sin(\Omega t):=t\sum_{n=0}^\infty
    \frac{(-1)^n}{(2n+1)!}\;(t^2\Omega^2)^n.
    \label{cossin}
    \ee
In this section we will study the structure of the solution
(\ref{E=1}) for the cases that the permittivity and permeability
tensors, $\Ep$ and $\MU$, are given by constant non-Hermitian
$3\times 3$ matrices. We achieve this by performing a Fourier
transform.

In the Fourier basis the curl operator $\fD$ is represented by the
antisymmetric Hermitian matrix:
    \begin{equation}
    \fD=\left(
    \begin{array}{ccc}
    0 & ik_{3} & -ik_{2} \\
    -ik_{3} & 0 & ik_{1} \\
    ik_{2} & -ik_{1} & 0 %
    \end{array}%
    \right),~~~~~~~k_{1}, k_{2}, k_{3} \in \mathbb{R},
    \label{1}
    \end{equation}
while $\Ep^{-1}$ and $\MU^{-1}$ that appear in the expression
(\ref{Omega1}) for $\Omega^2$ are $3\times3$ complex
matrices.\footnote{Given a linear operator acting in $\cH$, its
representation in the Fourier basis, $\tilde L$, is related to its
representation in the position basis, $L$, according to: $\tilde
L\tilde{\vec F}(\vec k)=(2\pi)^{-3/2}\int_{\R^3} e^{-i\vec
k\cdot\vec x}L\vec F(\vec x)\, d\vec x^3$. Here $\vec F$ is an
arbitrary complex vector field having a Fourier transform
$\tilde{\vec F}$ defined by $\tilde{\vec F}(\vec
k):=(2\pi)^{-3/2}\int_{\R^3} e^{-i\vec k\cdot\vec x}\vec F(\vec x)
\,d\vec x^3$. Throughout this article we use the same symbol for the
representations of an operator in both Fourier and position bases to
simplify the notation. This convention does not lead to any
confusion, for we exclusively use the Fourier representation of the
relevant operators.} It is not difficult to compute $\Omega^2$ in
the Fourier basis. This gives rise to a $3\times3$ matrix involving
$k_i$ and the entries of $\Ep$ and $\MU$, i.e., a total of 39 real
parameters. Rather than giving this complicated expression, we will
examine the Jordan canonical form of $\Omega^2$. In fact, it is more
convenient to work with the dimensionless operator,
    \be
    \hat\Omega^2:=\omega_0^{-2}\Omega^2=
    \omega_0^{-2}\,\Ep^{-1}\fD\,\MU^{-1}\fD,
    \label{Omega}
    \ee
where
    \be
    \omega_0:=c\,|\vec k|=c\sqrt{k_1^2+k_2^2+k_3^2}\:,
    ~~~~~~~
    \vec k:=\left(\begin{array}{c}k_1\\k_2\\k_3\end{array}\right).
    \label{omega-zero-k}
    \ee
In terms of $\hat\Omega^2$, (\ref{E=1}) and (\ref{cossin}) read
    \bea
    &&\vec E(\vec r,t)=\cos(\hat\Omega\,\omega_0t)\vec E_0(\vec r)+
    \omega_0^{-1}\hat\Omega^{-1}\sin(\hat\Omega\,\omega_0t)\dot{\vec E}_0(\vec r),
    \label{E=}\\
    &&\cos(\hat\Omega\,\omega_0t):=\sum_{n=0}^\infty
    \frac{(-1)^n}{(2n)!}\;(\omega_0^2t^2\hat\Omega^2)^n,
    \label{cos}\\
    &&\hat\Omega^{-1}\sin(\hat\Omega\,\omega_0t):=\omega_0t\sum_{n=0}^\infty
    \frac{(-1)^n}{(2n+1)!}\;(\omega_0^2t^2\hat\Omega^2)^n.
    \label{sin}
    \eea

A straightforward consequence of (\ref{1}) is $\fD \vec k=\vec 0$.
Therefore, $\fD$ has a zero eigenvalue and owing to its symmetry a
pair of real eigenvalues with opposite sign, namely $\pm|\vec k|$.
In view of (\ref{Omega}), this implies that $\hat\Omega^2$ will also
have a zero eigenvalue. Hence, its Jordan canonical form
$J_{\Omega^{2}}$ must have one of the following two forms.
    \bea
    {\rm Case~1}:&&
    J_{\Omega^{2}}=\left(
    \begin{array}{ccc}
    0 & 0 & 0 \\
    0 & \lambda_{-} & 0 \\
    0 & 0 & \lambda_{+} \end{array}
    \right),~~~~~\lambda_{\pm}\in\mathbb{C},
    \label{Case1}\\
    {\rm Case~2}:&&
    J_{\Omega^{2}}=\left(
    \begin{array}{ccc}
    0 & 0 & 0 \\
    0 & \lambda & 1 \\
    0 & 0 & \lambda
    \end{array}
    \right),~~~~~~~ \lambda \in
    \mathbb{C}.
    \label{Case2}
    \eea
These correspond to diagonalizable and non-diagonalizable cases,
respectively.

Let $S$ be an invertible matrix fulfilling
$\hat\Omega^2=S^{-1}J_{\Omega^{2}}S$. Then, in view of (\ref{cos})
and (\ref{sin}),
    \begin{equation}
    \cos(\omega_0t\hat\Omega)=
    S^{-1}\left(\sum_{n=0}^{\infty}\frac{(-1)^{n}}{(2n)!}
    \left(\omega_0^2t^{2}J_{\Omega^{2}}\right)^{n}\right)S,
    \label{ze1}
    \end{equation}
    \begin{equation}
    \hat\Omega^{-1}\sin(\omega_0t\hat\Omega)=
    S^{-1}\left(\omega_0t\sum_{n=0}^{\infty}\frac{(-1)^{n}}{(2n+1)!}
    \left(\omega_0^2t^{2}J_{\Omega^{2}}\right)^{n}\right)S.
    \label{ze2}
    \end{equation}
In order to simplify these equations we consider Cases 1 and 2
separately.

\begin{itemize}
    \item[] \underline{\textbf{Case 1} ($\hat\Omega^2$ is diagonalizable)}:
    In this case,
    \bea
    (J_{\Omega^{2}})^{n}=\left(
    \begin{array}{ccc}
    0 & 0 & 0 \\
    0 & \lambda_{-}^{n} & 0 \\
    0 & 0 & \lambda_{+}^{n}
    \end{array}
    \right)~~~~~~~\mbox{for all $n\in\Z^+$}.\nn
    \eea
Inserting this relation in (\ref{ze1}) and (\ref{ze2}) yields
    \bea
    &&\cos(\omega_0t\hat\Omega)=S^{-1}\left(
    \begin{array}{ccc}
    1 & 0 & 0 \\
    0 & \cos(\sqrt{\lambda_{-}}\,\omega_0t) & 0 \\
    0 & 0 & \cos(\sqrt{\lambda_{+}}\,\omega_0t) %
    \end{array}%
    \right)S,
    \label{sol}\\
    &&
    \hat\Omega^{-1}\sin(\omega_0t\hat\Omega)= S^{-1}\left(
    \begin{array}{ccc}
    0 & 0 & 0 \\
    0 & \frac{1}{\sqrt{\lambda_{-}}}
    \sin(\sqrt{\lambda_{-}}\,\omega_0t) & 0 \\
    0 & 0 & \frac{1}{\sqrt{\lambda_{+}}}
    \sin(\sqrt{\lambda_{+}}\,\omega_0t) %
    \end{array}%
    \right)S \label{sol1}
    \eea
Note that the time-harmonic solutions of the wave
equation~(\ref{E=}) are the eigenvectors of $\hat\Omega^2$,
\cite{epl-2008,p93}. Hence, they are given by
    \be
    \vec E^{(a)}_\pm(\vec r,t)=\int_{\R^3} d^3k\:\left[
    \sF_\pm(\vec k)e^{i(\vec k\cdot\vec
    r-\sqrt{\lambda_\pm}\,\omega_0t)}+
    \sG_\pm(\vec k)e^{i(\vec k\cdot\vec
    r+\sqrt{\lambda_\pm}\,\omega_0t)}\right] S^{-1}
    \hat e_a,~~~~~~a=1,2,
    \label{TD1}
    \ee
where $\sF_\pm$ and $\sG_\pm$ are complex-valued functions, and
    \[ \hat e_1=\left(\begin{array}{c}
    0 \\ 1 \\ 0\end{array}\right),~~~~~~
    \hat e_2=\left(\begin{array}{c}
    0 \\ 0 \\ 1\end{array}\right).\]
Clearly, for $a=1$ and $a=2$, $S^{-1} \hat e_a$ is the second and
the third column of the matrix $S^{-1}$, respectively. Furthermore,
the matrix $S$ and the eigenvalues $\lambda_\pm$ that enter
(\ref{TD1}) are, in general, functions of $\vec k$.

    \item[] \underline{\textbf{Case 2} ($\hat\Omega^2$ is
    non-diagonalizable)}: In this case,
    \begin{equation}
    (J_{\Omega^{2}})^{n}=\left(
    \begin{array}{ccc}
    0 & 0 & 0 \\
    0 & \lambda^{n} & n\lambda^{n-1} \\
    0 & 0 & \lambda^{n} %
    \end{array}%
    \right) ,~~~~~~~\mbox{for all $n\in\Z^+$},
    \label{22}
    \end{equation}
and we find
    \bea
    &&\cos(\omega_0t\hat\Omega)=S^{-1}\left(
    \begin{array}{ccc}
    1 & 0 & 0 \\
    0 & \cos(\sqrt{\lambda}\,\omega_0t) &
     -\frac{\omega_0t}{2\sqrt{\lambda}}\sin(\sqrt{\lambda}\,\omega_0t)
    \\
    0 & 0 & \cos(\sqrt{\lambda}\,\omega_0t) %
    \end{array}%
    \right)S,
    \label{sol0}\\
    && \hat\Omega^{-1}\sin(\omega_0t\hat\Omega)=S^{-1}\left(
    \begin{array}{ccc}
    0 & 0 & 0 \\
    0 & \frac{1}{\sqrt{\lambda}}\sin(\sqrt{\lambda}\,\omega_0t) &
    \frac{1}{2\lambda}\left[\omega_0t\cos(\sqrt\lambda\,\omega_0t)-
    \frac{1}{\sqrt\lambda}\sin(\sqrt\lambda\,\omega_0t)\right]
    \\
    0 & 0 & \frac{1}{\sqrt{\lambda}}
    \sin(\sqrt{\lambda}\,\omega_0t) %
    \end{array}%
    \right)S.~~~~~~~~~~
    \label{sol2}
    \eea
Moreover, the time-harmonic solutions of the wave equation
(\ref{E=}) take the form
    \be
    \vec E_\pm(\vec r,t)=\int_{\R^3} d^3k\:\left[
    \sF(\vec k)e^{i(\vec k\cdot\vec
    r-\sqrt{\lambda}\,\omega_0t)}+
    \sG(\vec k)e^{i(\vec k\cdot\vec
    r+\sqrt{\lambda}\,\omega_0t)}\right] S^{-1}
    \hat e_2.
    \label{TD2}
    \ee
where $\sF$ and $\sG$ are complex-valued functions.

\end{itemize}

We close this section by revealing an implicit symmetry of the
operator $\Omega^2$. Here we will momentarily consider the
possibility that permittivity and permeability tensors are not
constant.

As we have seen, $\Omega^{2}$ admits a single zero eigenvalue. The
presence of this zero eigenvalue implies the existence of parts of
$\Ep^{-1}$ and $\MU^{-1}$ that have no influence on the propagating
electric field. In order to characterize these, we expand $\Ep^{-1}$
and $\MU^{-1}$ as follows:
    \bea
    \Ep^{-1}= \ep_{\!\!\!H_{0}}+\ep_{\!\!\!H_{1}}+
    \ep_{\!\!\!A_{0}} +\ep_{\!\!\!A_{1}} , \label{11b}\\
    \MU^{-1}= \Mu_{\!\!\!H_{0}}+\Mu_{\!\!\!H_{1}}+
    \Mu_{\!\!\!A_{0}}+\Mu_{\!\!\!A_{1}}. \label{11c}
    \eea
Here the subscripts $H$ and $A$ denote the ``Hermitian'' and
``anti-Hermitian'' parts of the tensors respectively,
    \begin{equation}
    \ep_{\!\!\!H_{1}}\mathfrak{D}= \ep_{\!\!\!A_{1}}\mathfrak{D}=
    \Mu_{\!\!\!H_{1}}\mathfrak{D}=\mathfrak{D}\Mu_{\!\!\!H_{1}}=
    \Mu_{\!\!\!A_{1}}\mathfrak{D}=\mathfrak{D}\Mu_{\!\!\!A_{1}}=\Zero,
    \label{11d}
    \end{equation}
and $\Zero$ stands for the $3\times 3$ zero matrix.

Because the null space of $\Omega^2$ is spanned by $\vec k$,
according to (\ref{11d}) we have
    \begin{equation}
    \ep_{\!\!\!H_{1}}=\alpha_\varepsilon(\vec k)\:\vec k\vec k^\dagger,~~~
    \ep_{\!\!\!A_{1}}=i\beta_\varepsilon(\vec k)\:\vec k\vec k^\dagger,~~~
    \Mu_{\!\!\!H_{1}}=\alpha_\mu(\vec k)\:\vec k\vec k^\dagger,~~~
    \Mu_{\!\!\!A_{1}}=i\beta_\mu(\vec k)\:\vec k\vec k^\dagger,
    \label{11e}
    \end{equation}
where $\alpha_\varepsilon,\beta_\varepsilon,\alpha_\mu,\beta_\mu$
are real-valued functions of $\vec k$, and
    \[\vec k\vec k^\dagger=\left(
    \begin{array}{ccc}
    k^{2}_{1} & k_{1}k_{2} & k_{1}k_{3} \\
    k_{1}k_{2} & k^{2}_{2} & k_{2}k_{3} \\
    k_{1}k_{3} & k_{2}k_{3} & k^{2}_{3} %
    \end{array}\right).\]
For the cases that
    \begin{equation}
    \Ep_{A_{0}}^{-1}=\MU^{-1}_{A_{0}}=\Zero \label{11f}
    \end{equation}
$\Omega^2$ is pseudo-Hermitian. If in addition $\ep_{\!\!\!H_{0}}$
is positive definite, then $\Omega^{2}$ is quasi-Hermitian. Note
that these hold irrespectively of the value of $\ep_{\!\!\!H_{1}},
\ep_{\!\!\!A_{1}},\Mu_{\!\!\!H_{1}}$, and $\Mu_{\!\!\!A_{1}}$.

The above observation seems to be relevant only for the material
with spatial dispersion \cite{AG}. For the non-dispersive material
that we consider in this article, $\Ep$ and $\MU$ are independent of
$\vec k$ and the above gauge freedom is effectively frozen (the
scalar functions
$\alpha_\varepsilon,\beta_\varepsilon,\alpha_\mu,\beta_\mu$ are
identically zero).

\section{Applications: Plane-Wave and Time-Harmonic Solutions}

In this section we will examine the application of our general
results in the study of three concrete models. This also allows for
extracting the physical meaning of pseudo-Hermiticity and
quasi-Hermiticity of the $\Omega^2$.

\begin{itemize}
\item[] \textbf{Example~1:} Consider the propagation of a plane wave
given by the initial conditions
    \be
    \vec E_0(x,y,z)={\cal E}\:\left(\begin{array}{c}
    \cos\varphi
    \\ \sin\varphi
    \\0
    \end{array}\right)\:e^{ik_3z},~~~~~~~
    \dot{\vec E}_0(x,y,z)=\vec 0,
    \label{ini-con}
    \ee
in a uniaxial medium with permittivity and permeability tensors
\cite{figotin}:
    \begin{equation}
    \Ep=\varepsilon_0\left(
    \begin{array}{ccc}
    \varepsilon_1+i\gamma_\varepsilon & i\alpha & 0 \\
    -i\alpha & \varepsilon_1+i\gamma_\varepsilon & 0 \\
    0 & 0 & \varepsilon_3
    \end{array}
    \right), ~~~~~
    \MU=\mu_0\left(\begin{array}{ccc}
    \mu_1+i\gamma_\mu & i\beta & 0 \\
    -i\beta & \mu_1+i\gamma_\mu & 0 \\
    0 & 0 & \mu_3
    \end{array}%
    \right), \label{e1}
    \end{equation}
where ${\cal E}\in\C$ and $\varphi\in[0,2\pi)$ determine the
amplitude and polarization of the initial wave, $\varepsilon_0$ and
$\mu_0$ are the permittivity and permeability of the vacuum, so that
$\varepsilon_0\mu_0=1/c^2$, and
$\varepsilon_1,\varepsilon_2,\mu_1,\mu_2,\gamma_\varepsilon,\gamma_\mu,
\alpha,\beta\in\R$ describe the electric and magnetic properties of
the medium.\footnote{In light of (\ref{max-E}), $\dot{\vec
E}_0(x,y,z)=\vec 0$ follows from ${\vec B}_0(x,y,z)=\vec 0$.}

A simple consequence of (\ref{ini-con}) is that in the calculation
of $\hat\Omega^2$ in the Fourier basis we can set
    \be
    k_1=k_2=0.
    \label{k12=zero}
    \ee
This leads to an enormous simplification of the Jordan decomposition
of $\hat\Omega^2$. In particular, $\hat\Omega^2$ is diagonalizable,
$\omega_0=c|k_3|$, and inserting (\ref{1}) and (\ref{e1}) in
(\ref{Omega}) and calculating the eigenvalues of the resulting
expression for $\hat\Omega^2$ we find
    \be
    \lambda_\pm=[(\varepsilon_1\pm\alpha+i \gamma_\varepsilon)
    (\mu_1\pm\beta+i \gamma_\mu)]^{-1}.
    \label{f-omega=}
    \ee
Furthermore, the similarity transformation that diagonalizes
$\hat\Omega^2$ is independent of the parameters of the system, for
we have
    \be
    S=\left(\begin{array}{ccc}
    0 & 0 & 2\\
    i & 1 & 0\\
    -i & 1 & 0\end{array}\right),~~~~~
    S^{-1}=\frac{1}{2}\left(\begin{array}{ccc}
    0 & -i & i\\
    0 & 1 & 1\\
    1 & 0 & 0\end{array}\right).
    \label{S=z}
    \ee

Substituting (\ref{f-omega=}) and (\ref{S=z}) in (\ref{sol}) and
using the resulting relation and (\ref{ini-con}) in (\ref{E=}), we
obtain the following expression for the propagating electric field
    \be
    \vec E(\vec r,t)=\vec E(z,t)={\cal
    E}\vec n_E(t)\:e^{ik_3 z},
    \label{prop-E}
    \ee
where
    \be
    \vec n_E(t):=\frac{1}{2}
    \left(\begin{array}{c}
    e^{-i\varphi}\cos(\sqrt{\lambda_-}\:\omega_0 t)+
    e^{i\varphi}\cos(\sqrt{\lambda_+}\:\omega_0 t)\\
    i[e^{-i\varphi}\cos(\sqrt{\lambda_-}\:\omega_0 t)-
    e^{i\varphi}\cos(\sqrt{\lambda_+}\:\omega_0 t)]\\
    0\end{array}\right).
    \label{NE=}
    \ee
The expression for the magnetic field can be obtained using
(\ref{B}). Setting $\vec B(\vec r,0)=\vec 0$, which in view of
(\ref{max-B}) is consistent with the second equation in
(\ref{ini-con}), and using (\ref{prop-E}) and (\ref{B}), we find
    \be
    \vec B(\vec r,t)=i\,c^{-1}{\cal E}\,\vec n_B(t)\,e^{ik_3 z},
    \label{B=}
    \ee
where
    \be
    \vec n_B(t):=\frac{1}{2}
    \left(\begin{array}{c}
    i\left(\frac{e^{-i\varphi}\sin(\sqrt{\lambda_-}\:\omega_0
    t)}{\sqrt{\lambda_-}}-\frac{e^{i\varphi}\sin(\sqrt{\lambda_+}\:\omega_0
    t)}{\sqrt{\lambda_+}}\right)\\
    -\left(\frac{e^{-i\varphi}\sin(\sqrt{\lambda_-}\:\omega_0
    t)}{\sqrt{\lambda_-}}+\frac{e^{i\varphi}\sin(\sqrt{\lambda_+}\:\omega_0
    t)}{\sqrt{\lambda_+}}\right)\\
    0\end{array}\right).
    \label{NB=}
    \ee

It is remarkable that for the cases where $\Omega^2$ is a
pseudo-Hermitian operator but not quasi-Hermitian, so that
$\lambda_+=\lambda_-^*$, both the vectors $\vec n_E(t)$  and $\vec
n_B(t)$ stay real. Therefore, for the model we consider, the
propagating wave does not undergo a phase shift\footnote{According
to (\ref{prop-E}), $\vec E(\vec r,t)$ is given by a real
vector-valued function of $\vec x$ and $t$, namely $\vec n_E(\vec
x,t)$ times a complex scalar function of $\vec x$, i.e., ${\cal
E}e^{ik_3x}$. Hence we can associate a total phase factor to $\vec
E(\vec r,t)$, namely ${\cal E}e^{ik_3x}/|{\cal E}|$ which does not
change in time. In view of (\ref{B=}), the same is true for $\vec
B(\vec r,t)$.}; \emph{the pseudo-Hermiticity of $\Omega^2$ implies
phase conservation}. In light of (\ref{f-omega=}),
$\lambda_+=\lambda_-^*$ and $\Omega^2$ is pseudo-Hermitian (but not
quasi-Hermitian) if and only if $\mu_1\alpha \neq 0$ and
    \be
    \varepsilon_1\beta+\mu_1\alpha=0,~~~~
    \varepsilon_1\gamma_\mu+\mu_1\gamma_\varepsilon=0.
    \label{ph-condi}
    \ee

In view of (\ref{f-omega=}), these relations imply
    \be
    \lambda_\pm=\frac{\varepsilon_1}{\mu_1}\left[(\varepsilon_1^2+
    \gamma_\varepsilon^2-\alpha^2\mp i\alpha\gamma_\epsilon
    \right]^{-1}.
    \label{f-ph}
    \ee

Next, we derive the necessary and sufficient conditions for the
quasi-Hermiticity of $\Omega^2$ (equivalently $\hat\Omega^2$).
Because $\hat\Omega^2$ is diagonalizable, the latter is equivalent
to the reality of the eigenvalues $\lambda_\pm$ of $\hat\Omega^2$.
According to (\ref{f-omega=}), $\lambda_\pm$ are real and $\Omega^2$
is quasi-Hermitian if and only if either
    \be
    \gamma_\epsilon=\gamma_\mu=0,
    \label{q-condi1}
    \ee
or the following two conditions hold:
    \be
    \varepsilon_1\beta-\mu_1\alpha=0,~~~~
    \varepsilon_1\gamma_\mu+\mu_1\gamma_\varepsilon=0.
    \label{q-condi}
    \ee
Condition~(\ref{q-condi1}) corresponds to lossless media where $\Ep$
and $\MU$ are Hermitian. In this case $\Omega^2$ is known to be
quasi-Hermitian \cite{epl-2008}. Conditions~(\ref{q-condi}) do not
require Hermiticity of either of $\Ep$ and $\MU$, yet they imply
quasi-Hermiticity of $\Omega^2$. This is quite remarkable, because
whenever (\ref{q-condi}) holds,
    \be
    \lambda_\pm=\frac{\varepsilon_1}{\mu_1}\left[(\varepsilon_1\pm\alpha)^2+
    \gamma_{\epsilon}^2\right]^{-1}.
    \ee
Therefore, $\lambda_\pm$ are real and positive. According to
(\ref{prop-E}) -- (\ref{NB=}), this means that the amplitude of both
the electric and magnetic fields are bounded biperiodic functions of
time; \emph{quasi-Hermiticity of $\Omega^2$ confines the amplitude
of the propagating electric and magnetic fields to a bounded region
in the $x$-$y$ plane that does not shrink in time}.

In view of (\ref{ph-condi}) and (\ref{q-condi}), a necessary
condition for the pseudo-Hermiticity of $\Omega^2$ is the second
equation appearing in both (\ref{ph-condi}) and (\ref{q-condi}). For
the usual material where $\mu_1$ and $\varepsilon_1$ are positive,
this equation implies that $\gamma_\varepsilon$ and $\gamma_\mu$
have opposite sign. Therefore, to maintain the pseudo-Hermiticity of
$\Omega^2$ the material must simultaneously display loss with
respect to the electric properties and gain with respect to the
magnetic properties or vice versa.

As seen from (\ref{NE=}) and (\ref{NB=}), the propagating plane wave
determined by the initial conditions (\ref{ini-con}) is not a
time-harmonic solution of the Maxwell equations. It is rather a
superposition of two time-harmonic plane-wave solutions. In general,
the time-harmonic right-going, plane-wave solutions propagating
along the $z$-axis has the following form.
    \bea
    \vec
    E^{(1)}_\pm(z,t)&=&
    A^{(1)}_\pm
    \left(\begin{array}{c}
    1 \\ i \\0\end{array}\right) e^{i(k_3z-\sqrt{\lambda_\pm}\:\omega_0t)},~~~~~
    E^{(2)}_\pm(z,t)= A^{(2)}_\pm \left(\begin{array}{c}
    1 \\ -i \\0\end{array}\right)
    e^{i(k_3z-\sqrt{\lambda_\pm}\:\omega_0t)},~~~~
    \label{DH-PW}
    \eea
where $A^{(1)},A^{(2)}\in\C$. These solutions are clearly circularly
polarized. For the cases that $\Omega^2$ is quasi-Hermitian, i.e.,
either (\ref{q-condi1}) or (\ref{q-condi}) holds, (\ref{DH-PW}) are
periodic solutions with a constant amplitude. For the cases that
$\Omega^2$ is pseudo-Hermitian but not quasi-Hermitian, either $\vec
E^{(a)}_+$ or $\vec E^{(a)}_-$ is an exponentially decaying solution
while the other is an exponentially growing solution.

\item[] \textbf{Example~2:} Consider the propagation of electromagnetic
waves in a medium with complex symmetric permittivity and
permeability tensors \cite{ytk}:
    \begin{equation}
    \Ep=\varepsilon_0\La,~~~~\MU=\mu_0\La,~~~~
    \La:=\left(
    \begin{array}{ccc}
    \fa & \fg & \fu \\
    \fg & \fb & \fh \\
    \fu & \fh & \fc
    \end{array}
    \right),
    \label{non-diag}
    \end{equation}
where $\fa,\fb,\fc,\fg,\fh,\fu\in\C$. In this case, $\hat\Omega^2$
is a diagonalizable operator with a single nonzero eigenvalue,
namely
    \be
    \lambda_\pm=\lambda_0:=
    \frac{\fa\,k_1^2+\fn\, k_2^3+\fc\,k_3^2+2(\fg\,k_1k_2+
    \fh\,k_2k_3+\fu\,k_1k_3)}{|\vec k|^2[
    \fa\fb\fc+2\fg\fh\fu-(\fa\fh^2+\fb\fu^2+\fc\fg^2)]}.
    \label{eg2-eg-va}
    \ee
A direct consequence of this is the fact that pseudo-Hermiticity of
$\Omega^2$ coincides with its quasi-Hermiticity. The latter is
achieved by requiring that $\lambda_0$ be real.

We can easily diagonalize $\hat\Omega^2$ by setting
    \be
    S^{-1}=\left(\begin{array}{ccc}
    \frac{k_1}{k_3}& -\frac{\fu\,k_1+\fh\,k_2+\fc\,k_3}{\fa\,k_1+\fg\,k_2+
    \fu\,k_3}&
    -\frac{\fg k_1+\fb k_2+\fh\,k_3}{\fa k_1+\fg k_2+\fu k_3}\\
    \frac{k_2}{k_3}& 0 & 1 \\
    1 & 1 & 0\end{array}\right).
    \label{eg2-S}
    \ee

In the following we derive the expression for the time-harmonic
plane-wave solutions of Maxwell's equations for this case. In view
of (\ref{TD1}) and (\ref{eg2-S}), these are linear combinations of
    \bea
    \vec E^{(1)}(\vec r,t)&=&\frac{1}{|\vec k|}
     \left(\begin{array}{c}
    -(\fu k_1+\fh k_2+\fc k_3) \\ 0 \\\fa k_1+\fg k_2+\fu k_3
    \end{array}\right)
    (A^{(1+)} e^{i(\vec k\cdot\vec r-\sqrt{\lambda_0}\,\omega_0t)}+
    A^{(1-)} e^{i(\vec k\cdot\vec r+\sqrt{\lambda_0}\,\omega_0t)}),
    \label{eg2-DH-PW-1}\\
    \vec E^{(2)}(\vec r,t)&=& \frac{1}{|\vec k|}\left(\begin{array}{c}
    -(\fg k_1+\fb k_2+\fh k_3)\\ \fa k_1+\fg k_2+\fu k_3\\0\end{array}\right)
    (A^{(2+)} e^{i(\vec k\cdot\vec r-\sqrt{\lambda_0}\,\omega_0t)}+
    A^{(2-)} e^{i(\vec k\cdot\vec r+\sqrt{\lambda_0}\,\omega_0t)}),~~~~
    \label{eg2-DH-PW-2}
    \eea
where $A^{(1\pm)}$ and $A^{(2\pm)}$ are possibly $\vec k$-dependent
complex coefficients.

Again whenever $\Omega^2$ is quasi-Hermitian one obtains periodic
time-harmonic solution. Otherwise, depending on the imaginary part
of $\sqrt\lambda_0$ one may have an exponentially decaying or
growing solution. The latter is clearly dependent on the magnitude
and direction of $\vec k$. For example, for the special case of a
plane wave propagating along the positive $z$-axis, we have
    \be
    \vec E^{(1)}(z,t)=A^{(1+)}
     \left(\begin{array}{c}
    - \fc  \\ 0 \\ \fu
    \end{array}\right) e^{i(k_3z-\sqrt{\lambda_0}\:\omega_0t)},
    ~~~~~~
    \vec E^{(2)}(z,t)= A^{(2+)}
    \left(\begin{array}{c}
    - \fh \\ \fu\\0\end{array}\right)
     e^{i( k_3z-\sqrt{\lambda_0}\:\omega_0t)},
    \label{eg2-DH-PW-4}
    \ee
where $\omega_0=c\,k_3$ and
    \be
    \lambda_0=\left[\fa\fb+\frac{2\fg\fh\fu}{\fc}-
    \left(\frac{\fa\fh^2+\fb\fu^2}{\fc}+\fg^2\right)\right]^{-1}.
    \ee
For the special case, $\fa=\fb=(1+\fu^2)/\fc$, $\fg=\fu^2/\fc$,
$\fh=\fu$, that is considered in \cite{ytk}, we have
$\lambda_0=\fc^2$. Therefore, $\Omega^2$ is quasi-Hermitian and the
plane wave solutions (\ref{eg2-DH-PW-4}) do not decay in time
provided that $\fc$ is real. They are exponentially decaying
(growing) solutions, for $\IM(\fc)>0$ ($\IM(\fc)<0$).

\item[] \textbf{Example~3:} Consider a time-harmonic plane wave
propagating along the positive $z$-axis in a medium with complex
symmetric permittivity and permeability tensors:
    \begin{equation}
    \Ep=\varepsilon_0\left(
    \begin{array}{ccc}
    \ff-i\fg & \fg & 0 \\
    \fg & \ff+i\fg & 0 \\
    0 & 0 & 1
    \end{array}
    \right),~~~~~\MU=\mu_0\One,
    \label{eq3-non-diag}
    \end{equation}
where $\ff$ and $\fg$ are nonzero (possibly) complex parameters. In
this case, $k_1=k_2=0$, $\omega_0=ck_3$, and we can easily show that
$\hat\Omega^2$ is a non-diagonalizable operator, i.e., it is an
example of Case~2 of Section~2. Moreover, we have
    \be
    \lambda=\ff,~~~~~
    S=\left(\begin{array}{ccc}
    0 & 0 & 1\\
    0 & 1 & 0\\
    \fg & i\,\fg & 0\end{array}\right),~~~~~
    S^{-1}=\left(\begin{array}{ccc}
    0 & -i & \fg^{-1}\\
    0 & 1 & 0\\
    1 & 0 & 0\end{array}\right).
    \label{eg3-S}
    \ee
According to (\ref{TD2}) the propagating electric field is given by
    \be
    \vec E (z,t)=A^+\left(\begin{array}{c}
    1\\i\\0\end{array}\right)
    e^{i(k_3z-\sqrt{\ff}\:\omega_0 t)},
    \label{eg3-TD2}
    \ee
where $A^+\in\C$. Again $\Omega^2$ is pseudo-Hermitian provided that
$\ff$ is real. In this case, (\ref{eg3-TD2}) is periodic in time.
Otherwise, its amplitude is an exponentially decreasing or
increasing function of time. These correspond to $\IM(\sqrt\ff)>0$
and $\IM(\sqrt\ff)<0$, respectively.

Another peculiarity of the model considered here is that the
expression (\ref{eg3-TD2}) for the time-harmonic plane-wave solution
does not involve $\fg$. We only require that $\fg$ takes a nonzero
value. Furthermore, (\ref{eg3-TD2}) describes a right-going
clockwise circularly polarized field. The non-diagonalizability of
$\Omega^2$ implies that there is no other  time-harmonic plane wave
solutions propagating along the positive $z$-axis that is linearly
independent of (\ref{eg3-TD2}). We may take this as an indication of
the ``left-handedness'' of the medium. This observation motivates
the solution of the Maxwell's equation associated with the following
initial conditions.
    \be
    \vec E_0 (\vec r)= {\cal E}\left(\begin{array}{c}
    1\\-i\\0\end{array}\right)
    e^{ik_3z},~~~~~~~~~~\vec B_0 (\vec r) =\vec 0.
    \label{eg3-ini-condi}
    \ee
The result reads
    \bea
    \vec E (\vec r,t)&=&\vec E(z,t)={\cal E}
    \left(\begin{array}{c}
    \cos(\sqrt\ff\,\omega_0t)+
    \frac{i\fg\,\omega_0t}{\sqrt\ff}\,\sin(\sqrt\ff\,\omega_0t)\\
    -i\cos(\sqrt\ff\,\omega_0t)-
    \frac{\fg\,\omega_0t}{\sqrt\ff}\,\sin(\sqrt\ff\,\omega_0t)
    \\0\end{array}\right)
    e^{ik_3z},
    \label{eg3-E}
    \\
    \vec B (\vec r,t)&=&\vec B(z,t)=
    c^{-1}{\cal E}
    \left(\begin{array}{c}
    1-\frac{i\fg\,\omega_0^2t^2}{2}\\
    -i(1+\frac{i\fg\,\omega_0^2t^2}{2})
    \\0\end{array}\right)
    \frac{\sin(\sqrt\ff\,\omega_0t)e^{ik_3z}}{\sqrt\ff},
    \label{eg3-B}
    \eea
where we have made use of (\ref{E=}), (\ref{Case2}), (\ref{sol0}),
(\ref{sol2}), and (\ref{B}). The appearance of the factors $\omega_0
t$ and $\omega_0^2 t^2$ as coefficients of the period functions of
$t$ in (\ref{eg3-E}) and (\ref{eg3-B}) is a clear indication that
these solutions are not time-harmonic.\footnote{Note that $\fg\neq
0$.} Indeed, even for the cases that $\ff$ is real, these solutions
are unbounded functions of time. This in turn means that one cannot
maintain right-handed circularly polarized initial waves of the form
(\ref{eg3-ini-condi}) in such a material.\footnote{There is not
reason to believe that such exotic material can exist in real life.}

\end{itemize}

\section{Concluding Remarks}

The information on the propagation of electromagnetic waves in a
stationary, possibly inhomogeneous, and anisotropic media is encoded
in a generally matrix-valued differential operator $\Omega^2$. In
the absence of loss and gain, this operator is quasi-Hermitian and
one can use the properties of quasi-Hermitian operator to obtain the
solution of the wave equation. In the presence of gain or loss,
there is no guarantee that this operator is diagonalizable or has
real eigenvalues.

In this article we examined the Jordan decomposition of this
operator for a class of anisotropic active media and showed how the
spectral properties of $\Omega^2$ might be used in the description
of the propagating waves. In particular, we examined the
time-harmonic and plane-wave solutions for three class of toy
models, derived explicit conditions on the permittivity and
permeability tensors that would render $\Omega^2$ pseudo-Hermitian
or quasi-Hermitian, and offered a physical interpretation of the
pseudo-Hermiticity and quasi-Hermiticity of $\Omega^2$ in terms of
the behavior of the propagating plane-waves solutions.

An interesting observation we made was that $\Omega^2$ might
actually be non-diagonalizable. We constructed an explicit model
with this property and demonstrated a surprising feature of this
model that could be interpreted as its left-handedness. Whether such
exotic material can exist in nature or be manufactured is a subject
of a separate investigation.

Our results may be extended in at least the following two main
directions. Firstly, one may try to generalize the method to
non-homogeneous isotropic media. This can  be achieved using the
methods of Fourier analysis. Secondly, one might attempt to include
the effects of dispersion following the ideas presented in
\cite{p93}.

\np

\section*{Acknowledgments} This work has
been supported by the Scientific and Technological Research Council
of Turkey (T\"UB\.{I}TAK) in the framework of the project no:
108T009, and by the Turkish Academy of Sciences (T\"UBA).

\ed
\begin{thebibliography}{99}

\bibitem{jackson} J.\ D.\ Jackson, {\em Classical Electrodynamics}
(Wiley \& Sons, New York, 1975).

\bibitem{epl-2008} A.~Mostafazadeh and F.~Loran, Europhys.\ Lett.~{\bf 81},
10007 (2008).

\bibitem{p1} A.~Mostafazadeh, J.\ Math.\ Phys.\ {\bf 43}, 205-214
(2002).

\bibitem{quasi} F.~G.~Scholtz, H.~B.~Geyer, and F.~J.~W.~Hahne,
Ann.\ Phys.\ (NY) {\bf 213} 74-101 (1992).

\bibitem{p2-p3} A.~Mostafazadeh, J.\ Math.\ Phys.\ {\bf 43},
2814-2816 (2002) and 3944-3951 (2002).

\bibitem{p93} A.~Mostafazadeh, Phys.\ Lett.\ A \textbf{374}, 1307-1310
(2010).

\bibitem{chew} W.~C.~Chew, {\em Waves and Fields in Inhomogeneous Media}
(IEEE Press, New York, 1995).

\bibitem{someda} C.~G.~Someda, {\em Electromagnetic Waves} (Chapman \& Hall,
London, 1998).

\bibitem{figotin} A. Figotin and I. Vitebskiy, Phys. Rev. B \textbf{77},
104421 (2008).

\bibitem{AG} V.~M.~Agranovich and V.~L.~Ginzburg, {\em Crystal
Optics with Spatial Dispersion, and Excitons} (Springer, Berlin,
1984).

\bibitem{ytk} T.V.Yioultsis, T.~D.~Tsiboukis, and E.~E.~Kriezis,
IEEE Transactions Magnetic \textbf{34}, 2732-2735 (1998);\\
N.~V.~Kantartzis, T.~V.~Yioultsis, T.~I.~Kosmanis, and
T.~D.~Tsiboukis, IEEE Transactions Magnetic \textbf{36}, 907-911
(2000).




\end{thebibliography}
